# ARTIFICIAL INTELLIGENCE IN SOCIAL SCIENCE: A STUDY BASED ON BIBLIOMETRICS ANALYSIS

Juan-Jose Prieto-Gutierrez
*Complutense University of Madrid*
*Spain*
*ORCID 0000-0002-1730-862*

Francisco Segado-Boj
*Complutense University of Madrid*
*Spain*
*ORCID 0000-0001-7750-3755*

Fabiana Da Silva França
*Complutense University of Madrid*
*Brazil*
*ORCID 0000-0001-8330-4621*

**Abstract:** *Artificial intelligence (AI) is gradually changing the planet. Data digitisation, computing infrastructure and machine learning are helping AI tools to spread across all sectors of society. This article presents the results of a bibliometric analysis of AI-related publications in the social sciences over the last ten years (2013-2022). Most of the historical publications are taken into consideration with the aim of identifying research relevance and trends in this field. The results indicate that more than 19,408 articles have been published, 85% from 2008 to 2022, showing that research in this field is increasing significantly year on year. Clear domains or disciplines of research related to AI within the social sciences can be grouped into sub-areas such as law and legal reasoning, education, economics, and ethics. The United States is the country that publishes the most (20%), followed by China (13%). The influence of AI on society is inevitable and the advances can generate great opportunities for innovation and new jobs, but in the medium term it is necessary to adequately face this transition, setting regulations and reviewing the challenges of ethics and responsibility.*

## INTRODUCTION

Today, artificial intelligence could be identified as a field of science and engineering concerned with the computational understanding of what is commonly referred to as intelligent behavior, and the creation of artefacts that exhibit such behavior (Shapiro, 1992).

The origins of artificial intelligence date back to the mid-19th century, and it has been a subject of constant reflection up to the present day ever since Ada Lovelace published the first algorithm for a machine in 1843 (Aiello, 2016) thanks to which she can be considered the first programmer, and, years later, Charles Babbage designed his Analytical Engine (Babbage, 1973).

Many writings by Homer, René Descartes, Ramon Llull or Jules Verne used robots, mechanical beings and animals or artificial creations and have served as inspiration for many researchers interested in artificial intelligence.

In 1950, Alan Turing laid the foundations of computer science and automatic decoding, motivating the development of artificial intelligence through his advances and positing the Turing machines or the Turing test as simple abstract computational devices designed to help investigate the scope and limitations of what can be computed (Turing, 1950). Thus, it was confirmed that the computer was the most promising means to implement a great dream: to create intelligence made by man but not reproduced through the propagation of the species (Buchanan, 2005) and innovations began to appear in different areas of engineering, mainly in the military.

Years later, during World War II, Professor Aiken built the first mechanical digital machines at Harvard, with the help of IBM (Cohen, 2000). They gave rise to the first large-scale, automatically sequenced, general-purpose, functional digital computer produced in the United States (Aiken, 1964), which made decisions that until then could only be made by humans. They were used in the early 1960s by large companies such as General Motors (Robot Surg, 2007). A few years later, Stanford University marked a new era by presenting the first robot, named Shakey, capable of analysing instructions (Nilsson, 1984).

Around this time, the term "artificial intelligence" appeared for the first time in connection with a project of several American institutions. The study proceeded on the basis of the conjecture that every aspect of intelligence learning can, in principle, be described so precisely that a machine can be created to simulate it. Since then, the term has stuck around (McCorduck, et al., 1977).

Most of the 1970s is known as the "AI winter", meaning that, after a period of reduced funding and interest, fewer significant developments have originated (Floridi, 2020). Although other "winters" also occurred in later decades (Hendler, J. (2008). It was not until the 2000s that fundamental advances in AI took place.

With the year 2000, many of the existing limitations were overcome, giving way to a spring where AI systems began to analyze complex algorithms and self-learning became widespread (Havenstein, 2005). For example, in 2007, IBM created an open domain question and answer system, called Watson, which competed with human participants and won first place on the TV show Jeopardy! in 2011 (Kaul, et al. 2020). This technology, called DeepQA, used language processing and various searches to analyze data on unstructured content to generate probable answers (Ferrucci, et al, 2013). There is no doubt that the development of mobile Internet brought more artificial intelligence application scenarios (Zhang & Chen, 2020). Today, thousands of AI applications are deeply embedded in the infrastructure of all industries and sectors: automotive industry with mainly driverless cars (Tubaro & Casilli, 2019); financial markets with machine learning or bigdata (Duan, et al., 2019); smart payment systems through facial and voice recognitions (Dijmărescu, et al., 2022); health through healthcare, pharmaceuticals or disease detection (Ravì, et al., 2017); smart homes, through home automation and sensing systems (Kassens-Noor, et al, 2021); in the teaching and learning environment with ChatGPT (Baidoo-Anu & Owusu Ansah, 2023), etc.

When it comes to knowing what has been published throughout the history of artificial intelligence in the area of bibliometrics, we have a very detailed perspective by speciality and under-represented by the main disciplines (STEM, Life Sciences, Social Sciences and A&H). Bibliometrics, used as an evaluation criterion (Álvarez-Betancourt and García-Silvante, 2014) to generate really useful quantitative and statistical analyses (Pritchard, 1969), uses various methods to estimate scientific production, the analysis of journals, their authors, the activity of countries, institutions and even collaboration between them.

Currently, analyses related to IA in the field of health and medicine stand out, where (Guo, et al., 2020) show a growth in research articles of 45.15 % between 2014 and 2019. In a broader study from 1977 to 2018, it is exposed that the number of publications related to cancer was the highest, followed by heart disease and stroke (Tran, et al., 2019). To a lesser extent, AI research has been published in the areas of engineering (Shukla, et al, 2019), e-commerce (Bawack, et al, 2022), education (Song & Wang, 2020), university management (Prieto-Gutierrez, 2023), maritime (Munim, et al, 2020), agri-food chains (Monteiro, & Barata, 2021) and finance (Goodell, et al., 2021).

Of the scarce global research, we found one, not current, where through the Science Citation Index database it is concluded that scientific production increases widely from the 1990s, that computer science and engineering were the most active subjects in AI and the United States was the country that published the most on the subject (Niu, et al, 2016).

Against this background, this study aims to provide a holistic view of AI research in the social science discipline over the last ten years by assessing the conceptual development, composition, intellectual structure and dynamics of the research field (Zupic, & Čater, 2015). There is no doubt that this analysis is an essential resource for those who are fascinated by AI.

## MATERIALS AND METHODS

For the bibliometric analysis, the core collection of the Scopus database (Elsevier, 2018) is used to search for all existing AI publications. This database was chosen over others such as Web of Science because it is one of the largest multidisciplinary databases of academic literature, has greater coverage and provides advanced search capabilities.

After preparing the query, the search was carried out on 25th May using the criteria TITLE, KEYWORD and ABSTRACT as shown in the query. Using Boolean logical operators, the following search was conducted. Search query text:

> TITLE-ABS-KEY ( "artificial intelligence" ) AND ( LIMIT-TO ( SUBJAREA , "SOCI" ) ) AND ( LIMIT-TO ( PUBYEAR , 2022 ) OR LIMIT-TO ( PUBYEAR , 2021 ) OR LIMIT-TO ( PUBYEAR , 2020 ) OR LIMIT-TO ( PUBYEAR , 2019 ) OR LIMIT-TO ( PUBYEAR , 2018 ) OR LIMIT-TO ( PUBYEAR , 2017 ) OR LIMIT-TO ( PUBYEAR , 2016 ) OR LIMIT-TO ( PUBYEAR , 2015 ) OR LIMIT-TO ( PUBYEAR , 2014 ) OR LIMIT-TO ( PUBYEAR , 2013 ) )

A total of 19,408 scientific documents obtained were used in the pre-established bibliometric analysis. In order to obtain the largest number of articles in a specific period, the articles were limited to the last ten years (the year 2023 was excluded as it was incomplete) and no document types, languages or media types were excluded.

Therefore, the bibliometric study includes grey literature, conference proceedings, or books/book chapters. Articles written in any language other than English are included.

The data exported for the analysis focus on publication characteristics, content or contribution to the research topic, where activity indicators such as year of publication, number of documents, journals, countries, authors, universities, and other indicators of impact on scientific productivity such as the H-index in journals are considered. As it handles bibliographic information from thousands of documents, the download has been limited to several blocks of 2,000 documents, in order to carry out a bibliometric mapping focused on keywords and relationships between authors. The structural analysis of the networks and their representation was carried out using the specific software, Pajek (Batagelj, & Mrvar, 1998).

The data were prepared for analysis using information in Microsoft Excel where, through a statistical analysis of the metadata identified, the annual growth trajectory is sought.

## RESULTS

Table 1 show the initial research results retrieve 19,408 publications from the Scopus database related to IA research in Social Sciences discipline. As shown in figure 1 they were divided into 11 document types for the period 2013-2022. Articles represent 49.68% with 9,642, followed by Conference Papers with 33.24%, both types of documents account for around than 83% of the total number of publications.

**Table 1.** Number and types of documents in artificial intelligence publications.

| Document type | Number of results |
|---|---|
| Article | 9.642 |
| Conference Paper | 6.453 |
| Book Chapter | 1.653 |
| Review | 840 |
| Book | 332 |
| Editorial | 174 |
| Note | 111 |
| Conference Review | 100 |
| Erratum | 41 |
| Data Paper | 23 |
| Letter | 19 |

## Annual Production of Published Papers

The scientific production obtained in the period of this research is presented by year in Figure 1. Two periods of development of AI research in the Social Sciences could be identified in terms of the number of articles published, according to the behavior of the curve.

The first period, from 2013 to 2018, shows the low introduction of AI, with a very low academic output (before this period it would be considered residual). The second period, from 2018 to 2022, shows an exponential explosion in output, with a growth from 1,000 papers per year to around 5,000 in 2022. Arguably, 85% of publications are produced from 2008 to the present, and the rate of growth is expected to continue.

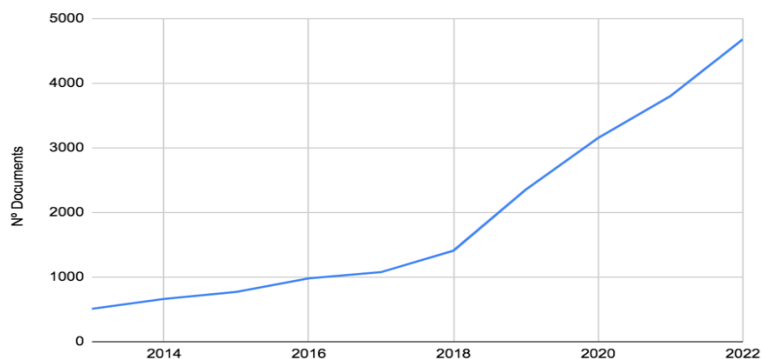

**Figure 1.** Annual publications on Scopus core database.

## Contribution by Country or Region

Authors from 150 different countries participated in the publication of the retrieved documents (644 documents with undefined country stand out). The top 10 active countries (see Table 2) participated in the publication of 13,833 documents (representing 70.91% of the total number of publications). The United States ranked first (3,625 documents with 19.58%) in terms of the number of publications. China is a close second (2,589 documents with 13.27%). The following positions are gradually declining in terms of presence.

**Table 2.** Top 10 Contribution by country or region 2013–2022.

|    | Country        | No. of contributions |
|----|----------------|----------------------|
| 1  | United States  | 3,625                |
| 2  | China          | 2,589                |
| 3  | United Kingdom | 1,616                |
| 4  | India          | 1,187                |
| 5  | Germany        | 1,025                |
| 6  | Italy          | 945                  |
| 7  | Australia      | 815                  |
| 8  | Spain          | 800                  |
| 9  | Canada         | 661                  |
| 10 | Russia         | 570                  |

Figure 2 shows the results of network analysis on the international collaborations between the high productive countries in AI in SSCC, from which we can see that the most internationally collaborative country is the United States, followed by China and United Kingdom, which are all developed countries. Some developing countries (such as India and Iran) present international collaboration, even with a very high academic production, as is the case of India.

In order to analyze the relationships of the different countries to publish on artificial intelligence in the area of Social Sciences, figure 2 illustrating a graph is shown.

For the elaboration of a network of co-authorships as adequate as possible, given the large number of countries that have published (150), a minimum threshold of 5 frequencies or publications has been set to appear in the maps. Approximately 30% would not meet this condition.

The analysis in Figure 2 shows 4 clusters according to the colours of the nodes. Two of them (yellow and green) are the most active and prominent. The cluster led by the United States exercises control over the rest of the countries, both in the number of publications and in the number of relations or contacts it maintains. It maintains strong links with countries on all continents, from Europe to Asia and the East. The relationship with China stands out, being very intense and the most numerous between two countries, which indicated by the thickest edge in the network, greater than that between the United States and the United Kingdom. China is placed with a yellow node because it is closely linked to the United States and because it has not created a unique and exclusive network. China's relations with neighboring countries such as Hong Kong, Singapore, India and Taiwan stand out.

The second most prevalent cluster is green and is led by United Kingdom, both in terms of number of publications (1,616) and relationships (10). Its largest and best link is the United States with 165 shared documents.

On the opposite side, and in an isolated and even withdrawn way, are the blue and red clusters, formed by the relationship of two countries (Portugal-Brazil and Saudi Arabia-India, respectively). The first two countries have no contact with the rest of the network. India, on the other hand, is this cluster's link to the East (China) and the West (the United States and the United Kingdom).

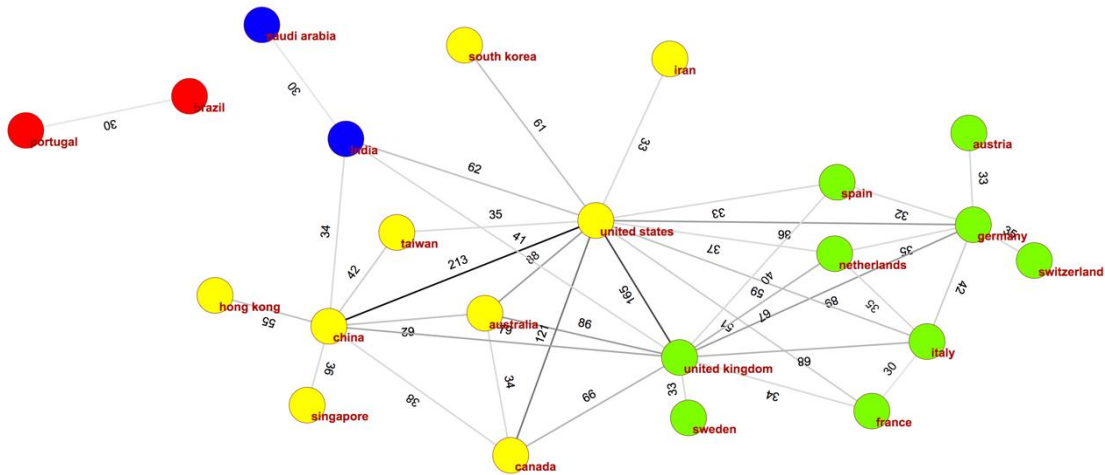

**Figure 2.** Network analysis of the international collaboration among the high productive countries.

## Co-occurrence Analysis of Keywords

In order to understand the knowledge components and knowledge structure of publications on artificial intelligence in the social sciences, the co-occurrence network of keywords is analysed by examining the links between keywords. It was decided to remove the term "Artificial Intelligence" because as the common term, it would be the huge centre of the network and would not allow to appreciate the relationships between terms.

Figure 3 illustrates this section with a network graph of keyword co-occurrences using the Pajek social network analysis tool. Authors select more than one keyword to define and guide the direction of their publication and are therefore essential indicators to analyse the lines and utilities of artificial intelligence in the area of Social Sciences.

Keywords are essential indicators that authors carefully select to define and represent the direction of their research article, encapsulating the basic theory, practical information and relevance of the article.

The network diagram not only shows the characteristics of a single concept related to artificial intelligence, but also describes the complex conceptual relationship between other concepts. Comparing keywords between AI-related concepts can also help to reveal their common associations and unique characteristics (related results are shown in Figure 3).

The most numerous nodes (keywords) are given in different colours. The links between the nodes are the co-occurrences of those two words. The number between the two is the weight of the link, so the numerical figure represents the activity. The higher the number the thicker the link shown should be.

Figure 3 groups the keywords into clusters to indicate which topics are the most common in artificial intelligence in the social sciences. The clusters are represented with different colours. There are 7 clusters which include at least 2 nodes.

In our analysis, there are 3 main themes and not all of them are related to each other. 3 large clusters appear with different colours as follows: White = machine learning, Orange = covid-19 and Red = big data.

It is supposed to be the core node of the cluster is dominated by the word "machine learning" and appears in white. Several machine learning terms share space in this cluster, such as "deep learning", "data mining" or "neural networks".

The cluster commanded by the word 'big data' appears in red, and exercises control over the other terms related to this activity, sharing space with "cloud computing", "data science" or "internet of things". These last nodes are, in turn, directly related to the main node "machine learning".

On the right-hand side of the figure are several independent and well-defined clusters that represent similarities in the research fields related to artificial intelligence. For example, in green are the nodes related to the field of education, in blue are the experiences on virtual and augmented reality that show information in the visual field of the user, aimed at museums, products and shopping, events, training, etc. Yellow represents the cluster that connects the nodes of accountability and transparency, setting the basis for ethics in artificial intelligence. Blue shows areas that add virtual elements to our real environment through "artificial intelligence", such as augmented and virtual reality. Pink shows the cluster related to automation and robotics, through artificial intelligence, of activities in industries or business processes.

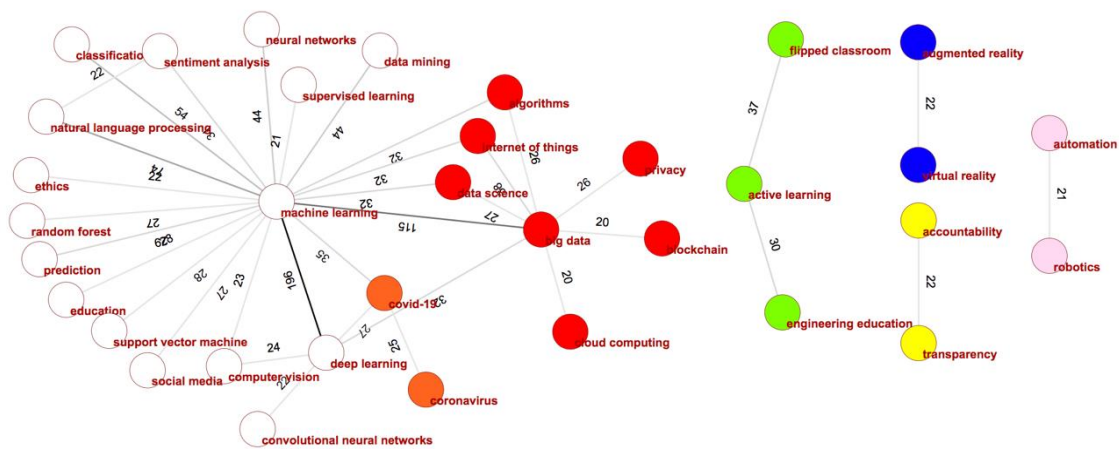

**Figure 3.** The conceptual network structure of keywords associated with the "artificial intelligence" concept.

## Leading Authors and Network of Co-authorships

Table 3 shows the 10 most prolific IA authors in the Social Sciences from 8 different countries: United Kingdom is the country contributing the most authors to the ranking. The remaining countries have only one representative.

The author with the most published articles, in the given period, is Gwojen Hwang, from the National Taiwan University of Science and Technology with 22 publications. Hwang started publishing on AI in 2020 in a remarkable way. The second author is the British Trevor Bench-Capon, from the Department of Computer Science at the University of Liverpool, who started publishing on IA in 1987.

**Table 3.** Top 10 contributors and their productivity for 2013–2022.

|    | Authorship pattern | Affiliation country | No. of contributions |
|----|--------------------|---------------------|----------------------|
| 1  | Hwang, G.J.        | Taiwan              | 22                   |
| 2  | Bench-Capon, T.    | United Kingdom      | 20                   |
| 3  | Dadios, E.P.       | Philippines         | 19                   |
| 4  | Hassanien, A.E.    | Egypt               | 18                   |
| 5  | Prakken, H.        | Netherlands         | 18                   |
| 6  | Walton, D.         | United Kingdom      | 17                   |
| 7  | Dale, R.           | USA                 | 16                   |
| 8  | Governatori, G.    | Australia           | 16                   |
| 9  | Atkinson, K.       | United Kingdom      | 14                   |
| 10 | Bizon, N.          | Rumania             | 14                   |

Within the analysis of research, it is necessary to understand scientific collaboration in order to discover the interaction between two or more scientists. In this case, co-authorship of a paper is an official statement of the participation of two or more authors.

In order to develop a network of co-authorships as appropriate as possible, given the large number of authors (41,860), a minimum threshold of 5 frequencies or publications has been set to appear in the maps.

Figure represents the visualization of the collaboration network of authors. The nodes, shown in colour, represent authors who are connected to each other.

Each node represents an author who published at least five papers in the selected period on artificial intelligence. Each line connects two authors and indicates the presence of at least one co-authored publication. The numbers associated with the lines show the activity between the two.

Figure 4 shows the network of authors publishing on artificial intelligence in the area of Social Sciences and includes 40 authors with more than 5 papers published during the period 2013-2022 and contains 19 relationships or components, 17 of which are outstanding co-authorships with between 6 and 9 collaborations each.

The visualization of collaborative networks in Figure 4 shows a very weak network of authors from different institutions in the field of macro-level research.

Even so, and according to the Scopus database, the most visible networks are the component formed by Atkinson K. and Bench-Capo (integrated to the University of Liverpool and experts in legislation) followed by the component Darwish A. and Hassanien A.E. integrated into the University of Helwan and Cairo, respectively, and experts in ethical aspects, impacts of climate change, etc. of AI) having collaborated bilaterally 13 and 10 times respectively and would have the highest degree of centrality and the highest strength of collaboration with other authors. The largest network consists of the Greeks, of Patras University, Perikos, Grivokostopoulou and Hatzilygeroudis with no more than 6 publications among the group (seeking advances in the area of learning and education through the use of artificial intelligence).

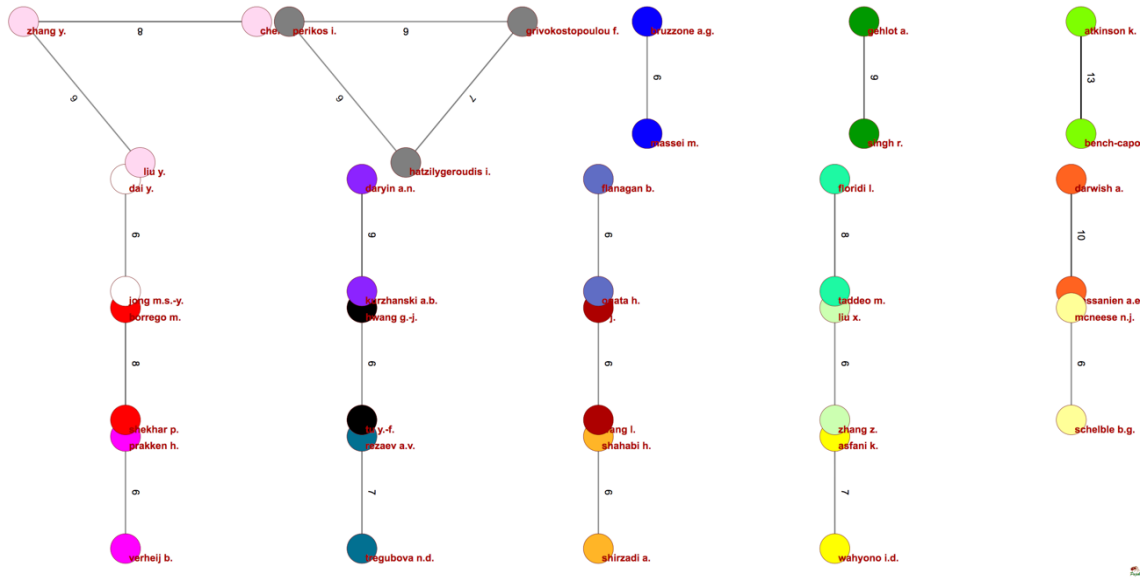

**Figure 4.** Authors cooperation network in Artificial Intelligence.

## CONCLUSIONS

Research in artificial intelligence in recent years has grown extensively, reaching everywhere and being used in our daily lives, in our jobs, in our leisure plans and in our homes, generating that many areas of activity can be transformed.

Given the tendency towards the dispersion or branches of application of artificial intelligence at present, this article has focused on the domain or discipline of social sciences. The social sciences are those that study the social and cultural aspects of human behaviour, which is why advances or transformations, as is the case with artificial intelligence, can lead to economic growth, build theories, generate changes in education, sociology, the application of laws and policies, etc. In short, it could improve the quality of life of many people and help to overcome global challenges, but as has been seen in an acceptable number of publications, aspects such as privacy or data protection can guarantee the ethical use of artificial intelligence.

Using the Scopus database, a search of titles, abstracts and keywords retrieved over 19,408 papers published in the last 10 years (2013-2022) focused on the Social Sciences, showing that artificial intelligence is a topic with a broad perspective global.

As has been seen in the analysis, IA has been a very active research field over the last 10 years, with 85% of all disclosures published in the last 5 years thanks to an exponential increase in publications. This fact demonstrates the gradual interest in AI in all fields and domains of Social Sciences in pursuit of the need to improve life development through continuous machine learning.

Pajek proves to be an effective tool for observing maps of relationships between authors, keywords and countries, revealing emerging trends in AI research, the intensity of keyword usage and visualizing the evolution of terms.

In terms of activity by country, China and the United States stand out for their dynamism and presence, both accounting for around 35% of the publications.

The United States, both in terms of volume and in terms of collaboration and contact, is the main focus of the publications network. It maintains relations with a large number of countries, either directly or indirectly, with collaboration with China standing out in an outstanding way.

The study of the network of collaboration between authors leads to the conclusion of the existence of small, well-defined, stable and secure groups over time. Two groups of three researchers stand out, with six or more publications together, whose origin is from Greece (in its totality) and the other trio is from China and Australia. The rest of the relationships are peer-to-peer and most of them are generated in their own institutions.

This shows that the most active groups are created and developed in their own organizations. Although, as mentioned above, there are dynamic collaborations between countries, these are still very scarce in terms of volume of activity.

The most active collaborative networks, Figure 4, focus on several fields. For example, the area related to legislation, due to the rapid advances and evolution of artificial intelligence, has generated research and subsequent publications in this field on the regulatory aspects and legal bases applicable to this new technology. In general, the research promotes innovation, ensures security and protects human rights.

Another active collaborative network, of researchers from Egypt, focuses on the impacts that IA can have on human activity. For example, in the recent covid-19 health crisis, they demonstrated the monitoring, surveillance, safety and recovery from disease using IA techniques, improving people's quality of life. Another topic of vital importance to society has focused on improving farming systems, water management, flood forecasting and climate change prediction through the power of data and artificial intelligence. They have even dared to forecast the GDP (Gross Domestic Product) of certain oil-exporting OPEC (Organisation of Petroleum Exporting Countries) members.

But the most active group, of Greek origin, has specialised in the area of teaching and learning by promoting innovative techniques. The field of education probably presents one of society's greatest challenges and, for this reason, research groups are active. In this case, the advances come from truly reforming experiences such as automatic learning; intelligent tutoring systems; the activation of virtual reality platforms to learn aspects of entrepreneurship; language learning; techniques that detect the state and behaviour of students through human-computer interaction; etc. In short, technological innovation in education, in terms of artificial intelligence, seeks to improve the quality of teaching processes in the institutions of each of the students by helping to fully personalise learning.

As shared in Figure 3, the largest percentage of publications deal with machine learning, covid and big data, but to a lesser extent, some research examines far-reaching aspects of the consequences and good use of artificial intelligence. Ethics is crucial because of its influence on automated decision-making that can affect individuals, organisations and society as a whole. On a planet, in which artificial intelligence is predicted to be booming, transparency is a decisive pillar of technological complexity and ethical implications. Against this backdrop of ethical dilemmas, standards and regulations from around the world are emerging to provide guidance and direction on ethical implementation.

With all the above, it can be confirmed that, the results confirm evident trends of growth in artificial intelligence studies and increases towards new challenges in a variety of research fields.

Even so, it would be appropriate to address the limitations of this bibliometric study. First, a limitation in the bibliometric analysis has been detected because, although Scopus is the most extensive database, it is possible that some important articles may have been missed as they are available in other databases, such as WoS or PubMed. It may even provide a varied understanding of other authors or contributors in the field under investigation and, in addition, may provide a new understanding of the main contributors in the field and reorganize the rankings among keywords, most influential researchers, and countries or regions. However, this research aims to reveal trends, not to offer a fixed formula or a ready-made recipe, and it fully complies.

## REFERENCES

Aiello, L. C. (2016). The multifaceted impact of Ada Lovelace in the digital age. *Artificial Intelligence*, *235*, 58-62. http://dx.doi.org/10.1016/j.artint.2016.02.003

Aiken, H. H. (1964). Proposed Automatic Calculating Machine. ed. and preface by A. C. Oettinger and T. C. Bartee. *IEEE Spectrum*, August, 62-9

Alvarez-Betancourt, Y., and Garcia-Silvente, M. (2014). An overview of iris recognition: a bibliometric analysis of the period 2000–2012. *Scientometrics, 101*, 2003-2033. http://dx.doi.org/10.1007/s11192-014-1336-1

Baidoo-Anu, D., and Owusu Ansah, L. (2023). *Education in the Era of Generative Artificial Intelligence (AI): Understanding the Potential Benefits of ChatGPT in Promoting Teaching and Learning*. http://dx.doi.org/10.2139/ssrn.4337484

Batagelj, V. and Mrvar, A. (1998). Pajek: Program for large network analysis. *Connections*, *21*(2), 47-57 https://webs.ucm.es/info/pecar/pajek.pdf

Bawack, R. E., Wamba, S. F., Carillo, K. D. A. and Akter, S. (2022). Artificial intelligence in E-Commerce: a bibliometric study and literature review. *Electronic markets. 32*(1), 297-338. 10.1007/s12525-022-00537-z

Buchanan, B. G. (2005). A (very) brief history of artificial intelligence. *Ai Magazine, 26*(4), 53-53. https://doi.org/10.1609/aimag.v26i4.1848

Charles Babbage. (1973). *Babbage's Calculating Engines. Charles Babbage Research Institute Reprint Series for the History of Computers*, Vol. 2. Los Angeles: Tomash Publishers.

Cohen, I.B. (1988). Babbage and Aiken. *Annals of the History of Computing*, *10*(3), 171-193, July-Sept., doi: 10.1109/MAHC.1988.10029.

Cohen, I. B. (2000). *Howard Aiken: Portrait of a computer pioneer*. MIT Press.

Dijmărescu, I., Iatagan, M., Hurloiu, I., Geamănu, M., Rusescu, C. and Dijmărescu, A. (2022). Neuromanagement decision making in facial recognition biometric authentication as a mobile payment technology in retail, restaurant, and hotel business models, *Oeconomia Copernicana*, 13(1), 225-250. https://doi.org/10.24136/oc.2022.007

Duan, Y., Edwards, J. S. and Dwivedi, Y. K. (2019). Artificial intelligence for decision making in the era of Big Data–evolution, challenges and research agenda, *International journal of information management, 48*, 63-71. https://doi.org/10.1016/j.ijinfomgt.2019.01.021

Elsevier. Scopus; Elsevier: Amsterdam, The Netherlands, 2018.

Ferrucci, D., Levas, A., Bagchi, S., Gondek, D. and Mueller, E. T. (2013). Watson: beyond jeopardy!, *Artificial Intelligence, 199*, 93-105. https://doi.org/10.1016/j.artint.2012.06.009

Floridi, L. (2020). AI and its new winter: From myths to realities, *Philosophy & Technology*, 33, 1-3. https://doi.org/10.1007/s13347-020-00396-6

Goodell, J. W., Kumar, S., Lim, W. M. and Pattnaik, D. (2021). Artificial intelligence and machine learning in finance: Identifying foundations, themes, and research clusters from bibliometric analysis, *Journal of Behavioral and Experimental Finance*, *32*(100577). https://doi.org/10.1016/j.jbef.2021.100577

Guo, Y., Hao, Z., Zhao, S., Gong, J. and Yang, F. (2020). Artificial intelligence in health care: bibliometric analysis. *Journal of Medical Internet Research*, *22*(7), e18228. https://doi.org/10.2196/18228

Havenstein, H. (2005). Spring comes to AI winter. *Computer World, 14*, 28-28.

Hendler, J. (2008). Avoiding another AI winter, *IEEE Intelligent Systems*, *23*(02), pp. 2-4.

Kassens-Noor, E., Wilson, M., Kotval-Karamchandani, Z., Cai, M. and Decaminada, T. (2021). Living with autonomy: Public perceptions of an AI-mediated future, *Journal of Planning Education and Research,* 0739456X20984529. https://doi.org/10.1177/0739456X20984529

Kaul, V., Enslin, S. and Gross, S. A. (2020). History of artificial intelligence in medicine. *Gastrointestinal endoscopy*, *92*(4), 807-812. 10.1016/j.gie.2020.06.040

McCorduck, P., Minsky, M., Selfridge, O. G. and Simon, H. A. (1977). August. *History of artificial intelligence*. IJCAI: 951-954.

Monteiro, J. and Barata, J. (2021). Artificial intelligence in extended agri-food supply chain: A short review based on bibliometric analysis. *Procedia Computer Scien*ce, 192, 3020-3029. https://doi.org/10.1016/j.procs.2021.09.074

Moran, M. E. (2007). Evolution of robotic arms. *Journal of robotic surgery,* 1(2), 103-111. https://doi.org/10.1007/s11701-006-0002-x

Munim, Z. H., Dushenko, M., Jimenez, V. J., Shakil, M. H. and Imset, M. (2020). Big data and artificial intelligence in the maritime industry: a bibliometric review and future research directions. *Maritime Policy & Management*, 47(5), 577-597. https://doi.org/10.1080/03088839.2020.1788731

Nilsson, N. J. (Ed.). (1984). *Shakey the robot.*

Niu, J., Tang, W., Xu, F., Zhou, X. and Song, Y. (2016). Global research on artificial intelligence from 1990–2014: Spatially-explicit bibliometric analysis. ISPRS International. *Journal of Geo-Information, 5*(5), 66. https://doi.org/10.3390/ijgi5050066

Prieto-Gutierrez, J.J. (2023). Los desafios de la inteligencia artificial en la gestión universitaria. *Diario Económico Cinco Días*, Grupo Prisa

Pritchard, A. (1969). Statistical bibliography or bibliometrics. *Journal of documentation, 25*, 348.

Ravì, D., Wong, C., Deligianni, F., Berthelot, M., Andreu-Perez, J., Lo, B. and Yang, G. Z. (2017). Deep learning for health informatics. *IEEE journal of biomedical and health informatics*, 21(1), 4-21. 10.1109/JBHI.2016.2636665

Shapiro, S. C. (1992). *Encyclopedia of artificial intelligence second edition.* New Jersey: A Wiley Interscience Publication.

Shukla, A. K., Janmaijaya, M., Abraham, A. and Muhuri, P. K. (2019). Engineering applications of artificial intelligence: A bibliometric analysis of 30 years (1988–2018). *Engineering Applications of Artificial Intelligence,* 85, 517-532. https://doi.org/10.1016/j.engappai.2019.06.010

Song, P., and Wang, X. (2020). A bibliometric analysis of worldwide educational artificial intelligence research development in recent twenty years. *Asia Pacific Education Review*, *21*, 473-486. https://doi.org/10.1007/s12564-020-09640-2

Tran, B. X., Vu, G. T., Ha, G. H., Vuong, Q. H., Ho, M. T., Vuong, T. T., ... and Ho, R. C. (2019). Global evolution of research in artificial intelligence in health and medicine: a bibliometric study. *Journal of clinical medicine, 8*(3), 360. https://doi.org/10.3390/jcm8030360

Tubaro, P., & Casilli, A. A. (2019). Micro-work, artificial intelligence and the automotive industry. *Journal of Industrial and Business Economics*, 46, 333-345. https://doi.org/10.1007/s40812-019-00121-1

Turing, (1950). Computing machinery and intelligence. *Mind*, *59*(236), 435–460.

Zhang, C. and Chen, Y. (2020). A review of research relevant to the emerging industry trends: Industry 4.0, IoT, blockchain, and business analytics. *Journal of Industrial Integration and Management*, 5(01), 165-180. https://doi.org/10.1142/S2424862219500192

Zupic, I., Čater, T. (2015). Bibliometric methods in management and organization. *Organizational research methods,* 18(3), 429-472. https://doi.org/10.1177/1094428114562629

## Authors' Note

This work was supported by funds from "Información, Biblioteca y Sociedad" research group, of the Complutense University of Madrid (Number 931763)

All correspondence should be addressed to
Juan-Jose Prieto-Gutierrez
Complutense University of Madrid
Ciudad Universitaria
jjpg@ucm.es